# Perfect Heat Rectification and Circulation with Nonreciprocal Radiative Surfaces in the Far Field

Sina Jafari Ghalekohneh[1] and Bo Zhao[1,2*]

[1]Department of Mechanical and Aerospace Engineering,

University of Houston, Houston, TX 77204, USA

[2]Department of Physics,

University of Houston, Houston, TX 77204, USA

[*]Corresponding author: bzhao8@uh.edu

**Abstract**

Controlling photon-mediated energy flow is central to the future of communications, thermal management, and energy harvesting technologies. Recent breakthroughs have revealed that many-body systems violating Lorentz reciprocity can sustain persistent photon heat current at thermal equilibrium, hinting at a new paradigm of heat flow akin to superconductivity. Yet, the behavior of such systems far from equilibrium remains largely unexplored. In this work, we uncover the rich physics of radiative heat transfer in nonequilibrium, far-field many-body systems composed of thermal emitters that break Lorentz reciprocity. We show that the total heat flow naturally decomposes into two distinct components: an equilibrium term, which generates a persistent circulating heat current within the system, and a nonequilibrium term, which governs energy exchange with the environment. Remarkably, while the internal persistent heat current is ever-present, the nonequilibrium contribution can be precisely engineered to achieve perfect heat rectification and circulation. Our results open a new route toward designing thermal systems with unprecedented control—unlocking the potential for lossless heat circulation and one-way thermal devices. This fundamentally shifts the landscape for next-generation thermal logic, energy conversion, and photonic heat engines.

Table of Content

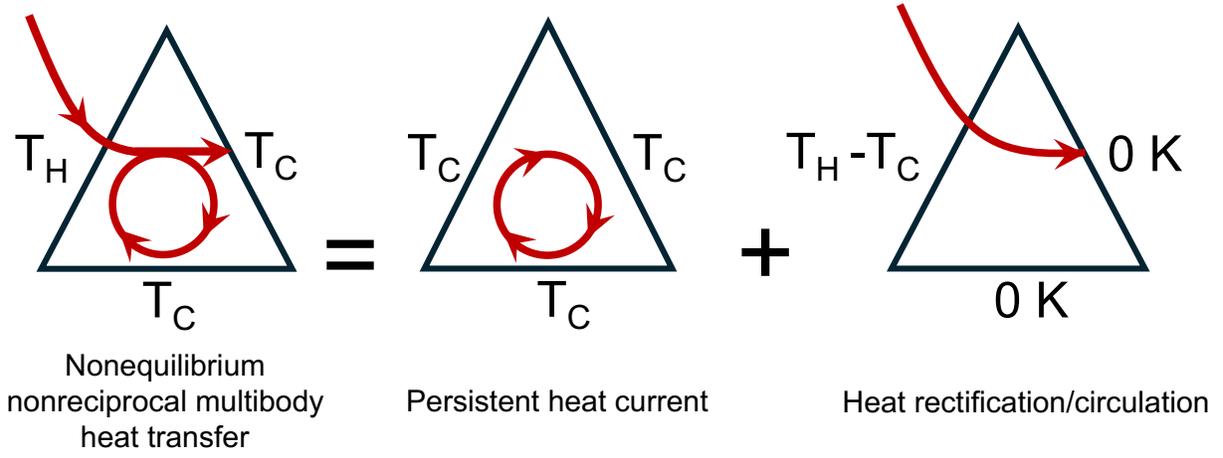

Nonequilibrium nonreciprocal multibody heat transfer = Persistent heat current + Heat rectification/circulation

# Introduction

The ability to control radiative heat flow often can unlock intriguing thermal and energy applications[1-5]. Heat rectifiers or diodes[6-9] can provide one-way heat transfer for thermal management and energy harvesting systems. Radiative heat circulation[10] allows the heat input to one body to be transferred only to the sequential body. These functionalities play critical roles in a wide spectrum of photon-mediated applications, including radiative energy harvesting[11-16] and thermal regulation[7,14,17-24].

Achieving perfect heat rectification and circulation is critical since any level of nonperfect rectification or circulation would imply backflow of heat flow, which contributes to parasitic heat loss and prevents energy harvesting systems and thermal management systems from reaching their thermodynamic limit[11-13,15]. Similar to optical counterparts[25], heat rectifiers and circulation require reciprocity breaking. However, as a critical difference, effective radiative heat transfer requires a broad wavelength and angular range. To achieve perfect heat rectification or circulation, reciprocity breaking is needed for all the $\omega$-$k$ channels for heat transport. Despite that it is known that reciprocity breaking can be achieved with nonlinearity[26-28], temporal modulation[29-32], or magnetic effects[33-36], the pathways of achieving perfect radiative heat rectification or circulation remain elusive.

Here, we study the possibility of achieving perfect heat rectification and circulation with nonequilibrium many-body systems consisting of thermal emitters that break Lorentz reciprocity. Research on this topic recently revealed many fascinating heat transfer phenomena. As an intriguing effect, it has been shown that when at thermal equilibrium, such systems can support a nonzero persistent heat current[17]. When at nonequilibrium, such systems could support thermal rectification[37,38], thermal hall effect[39], and thermal Corbino effect[40]. However, it is unclear whether perfect heat circulation can be achieved within a many-body nonreciprocal system. It also remains a question of how the transfer phenomena at equilibrium and nonequilibrium are related in these systems.

In this work, we report a pathway to achieve perfect heat rectification and circulation in a nonreciprocal many-body system. We show that the heat flow transport among the bodies when at nonequilibrium, in fact, consists of two parts: the equilibrium and nonequilibrium contributions. The equilibrium part results in a persistent heat current, whereas the nonequilibrium part contributes to the net heat transfer for each body. It is the nonequilibrium part that drives the energy transport with outside and can be controlled to achieve perfect heat rectification or circulation. Despite the reminiscence of the recently discovered persistent heat current[17] phenomena, we show that the nonzero persistent heat transport is only an accompanying effect and is fundamentally different from perfect heat circulation, which requires the system to operate out of equilibrium and has much more stringent requirements on the radiative properties of participating bodies. In contrast with previous studies that have been focused on near-field systems, we propose to use nonreciprocal bodies in the far field where the emissivity ($\varepsilon$) and absorptivity ($\alpha$) are no longer equal, i.e., violating the conventional Kirchhoff's law of thermal radiation[33,34,41]. Using a nonreciprocal ray-tracing technique, we show the pathway of achieving perfect rectification for all propagating wave channels through system configuration and directional nonreciprocal properties.

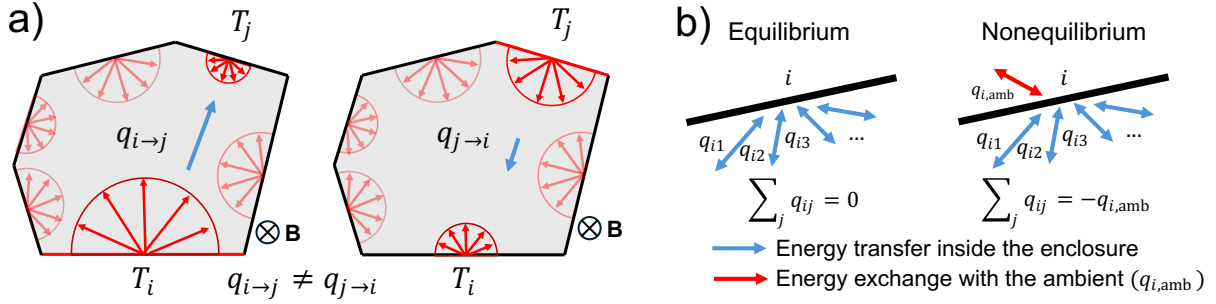

FIG. 1. Schematic of the system consisting of nonreciprocal surfaces. (a) The enclosure consisting of nonreciprocal emitters consisting of, for example, magneto-optical materials under an external magnetic field **B**. The arrows show the heat flows between two bodies, $i$ and $j$. (b) The energy balance schematic for body $i$ at equilibrium (left) and nonequilibrium (right) scenario.

# Methods

## A. Equilibrium system

We start by understanding the energy balance when the system is at equilibrium. We consider a general many-body system consisting of total $N$ bodies, as shown in Fig. 1(a). These surfaces form an enclosure. When body $i$ is emitting and $j$ is receiving, the spectral heat transfer rate from $i$ to $j$, $q_{i \to j}(\omega)$, is given by $q_{i \to j}(\omega) = N(\omega, T_i) G_{i \to j}(\omega)$, where $N(\omega, T_i)$ is the spectral energy density of the photons emitted by body $i$, and for a 2D system $N(\omega, T) = \frac{\hbar \omega^2}{2\pi c^2 \left( e^{\frac{\hbar \omega}{2\pi k_B T}} - 1 \right)}$, which is explained in detail in Supplement Material. $G_{i \to j}(\omega)$ is the energy transmission coefficient from body $i$ to $j$. The net spectral heat transfer rate between bodies $i$ and $j$ is[42-44]

$$q_{ij}(\omega) = q_{i \to j}(\omega) - q_{j \to i}(\omega) = N(\omega, T_i) G_{i \to j}(\omega) - N(\omega, T_j) G_{j \to i}(\omega). \tag{1}$$

In reciprocal systems, $G$ is symmetric upon switching source and receiver, yielding $G_{i\to j}(\omega) = G_{j\to i}(\omega)$ and thus the traditional formula[45] for spectral heat rate $q_{ij}(\omega) = G_{i\to j}(\omega)[N(\omega, T_i) - N(\omega, T_j)]$. Therefore, $q_{ij}(\omega) = 0$ when $T_i = T_j$. For a nonreciprocal system, $G_{i\to j}(\omega) \neq G_{j\to i}(\omega)$, causing a nonzero heat flow even when $T_i = T_j$, which is the basis of persistent heat current at thermal equilibrium[17].

The persistent heat current, however, does not involve heat transfer with the environment outside the system. When the many-body system is maintained at equilibrium, no heat is transferred from the ambient to the system. The net heat flow of each body should be zero based on the second law of thermodynamics, i.e., $\sum_j q_{ij} = 0$. Therefore, one can show that

$$q_{ij}(\omega) = -\sum_{k \neq j} q_{ik}(\omega). \tag{2}$$

In other words, the energy of the persistent heat current transferred between body $i$ and $j$ comes from the energy received by $i$ from all other bodies in the system. The energy extracted from outside the system by each body at equilibrium is zero. Therefore, the nonreciprocal equilibrium system will not be able to provide rectification or circulation functionalities to any heat input from outside the system, despite the fact that there is a nonzero heat flow circulating in the system.

B. Nonequilibrium system

When the system is at nonequilibrium, each body starts to have an energy exchange with the ambient. We assume the ambient is one single body. The heat transfer between body $i$ and the ambient is

$$q_{i,\text{amb}}(\omega) = -\sum_{k=1}^{N} q_{ik}(\omega), \tag{3}$$

where a positive $q_{i,\text{amb}}(\omega)$ indicates a heat output from body $i$ to the ambient. In fact, when each body receives or loses an amount of energy from or to other bodies, to keep its temperature constant, it needs to exchange this amount energy with the ambient. Without losing generality, we can consider the nonequilibrium scenario of a heat circulator, where one body has a higher temperature $T_h$ whereas the rest of the system remains at a lower temperature $T_c$. We note that all of the bodies exchange heat with the ambient, however the direction of this heat exchange is different. For bodies with lower temperature $T_c$ the heat exchange is from the body to the ambient while for the body with higher temperature $T_h$, the direction is the opposite. We set $T_i = T_h$ and consider the transfer between body $i$ and $j$. The heat flow, which follows Eq. (1), can be separated into two parts

$$q_{ij}(\omega) = N(\omega, T_c)[G_{i \to j}(\omega) - G_{j \to i}(\omega)] + [N(\omega, T_h) - N(\omega, T_c)]G_{i \to j}(\omega), \quad (4)$$

where the first term is the equilibrium contribution $q_{ij,\text{eq}}$, and the second term is the nonequilibrium contribution $q_{ij,\text{neq}}$. It is clear that $q_{ij,\text{eq}}$ is the term corresponding to the persistent heat current[17], and it is nonzero, as we are using nonreciprocal emitters, regardless of whether the system is at equilibrium or nonequilibrium. One can interpret this equilibrium part as a scenario when all bodies are kept in the same temperature $T_c$. Meanwhile, as in the equilibrium scenario, $q_{ij,\text{eq}}$ transfers heat only among the bodies within the system and does not involve the ambient, i.e., $\sum_j q_{ij,\text{eq}} = 0$. Therefore, the heat exchange between body $i$ and the ambient in Eq. (3), is contributed by the nonequilibrium part

$$q_{i,\text{amb}}(\omega) = -\sum_{k=1}^{N} q_{ik,\text{neq}}(\omega). \quad (5)$$

The above equations highlight the main theory of this study. Heat circulation and rectification require the heat input from outside the system to be circulated or rectified. Therefore, the figure of merit for rectification and circulator in nonreciprocal systems is defined based on the heat transfer rates by the nonequilibrium terms.

In an ideal circulator, one allows the temperature of one body in the system to increase, and a circulator should direct the heat flow to the sequential body. Following the desired heat flow direction, we assign each body an index with the energy from body *i* directed to body *j*, where *j* is the index of the sequential body.

Without losing generality, we consider an enclosure containing three surfaces, as shown in Fig. 2. The three surfaces form an equilateral triangle, and each surface has the same nonreciprocal radiative properties. The system, therefore, has a threefold rotational symmetry ($C_3$). In doing so, a perfect rectification between a pair of surfaces would naturally imply a perfect circulation among the three bodies, with the magnitude of heat flow that can be transferred between each pair of adjacent bodies the same. Therefore, we focus on achieving perfect rectification between two surfaces in the system, i.e., surfaces *i* and *j*.

### C. Heat rectification

In general, the heat rectification between *i* and *j* can be quantified by the rectification ratio $\gamma$ defined as

$$\gamma_{ij} = \frac{|q_{f,i} - q_{b,j}|}{q_{f,i} + q_{b,j}}, \tag{6}$$

where the forward-case heat flow, $q_{f,i}$, is the net total heat transfer rate from body *i* to *j* by the nonequilibrium term, $q_{ij,\text{neq}}$, when $T_i = T_h$ and body *j* and the rest of the system have a lower temperature $T_c$. The backward-case heat flow, $q_{b,j}$, is the net heat transfer rate from body *j* to *i* by the nonequilibrium terms, $q_{ji,\text{neq}}$, when $T_j = T_h$ and body *i* and the rest of the system have a lower temperature $T_c$. With Eq. (4), we have

$$\gamma_{ij} = \int \frac{|G_{i \to j}(\omega) - G_{j \to i}(\omega)|}{G_{i \to j}(\omega) + G_{j \to i}(\omega)} d\omega. \tag{7}$$

The energy transmission coefficient $G_{i \to j}(\omega)$ essentially describes the probability of photons emitted from surface *i* and captured by surface *j*. Therefore, a prefect rectification, i.e., $\gamma_{ij} =$ 1, would require either $G_{i\to j}(\omega)$ or $G_{j\to i}(\omega)$ to be zero. In the following, we focus on achieving a pathway that can accomplish $G_{i\to j}(\omega) = 0$.

Without losing generality, here we consider nonreciprocal emitters in the black-and-white symmetry group[46] since they can be easily accessible with pattern-free multilayer structures. Such materials include magneto-optical materials[24,34-36] or Weyl semimetals[18,47,48] which are already proven theoretically and experimentally to be able to show nonreciprocal behavior. The absorptivity and emissivity of these emitters can be related to their reflectivity as follows[20,33]:

$$\alpha(\theta) = 1 - r(\theta \to -\theta) \quad \text{and} \quad \varepsilon(\theta) = 1 - r(-\theta \to \theta) \qquad (8)$$

where $\theta$ is the polar angle with respect to the normal direction of the surface on the emitting side. Intuitively, we start from a nonreciprocal surface with an omnidirectional nonreciprocal property prescribed in Fig. 2(d). For all light incident from positive polar angles, regardless of their polarizations, $r(\theta \to -\theta) = 1$, and for all negative polar angle $-\theta$, $r(-\theta \to \theta) = 0$. Therefore, $\alpha = 0$ and $\varepsilon = 1$ for all positive $\theta$, and $\varepsilon = 0$ and $\alpha = 1$ for all negative polar angle $-\theta$.

To highlight the nonreciprocal effect, without losing generality, we examine the heat transfer confined within the *x-y* plane. We propose a ray tracing technique for general systems without reciprocity constraints, incorporating nonreciprocal radiative properties into ray emission and interaction steps. The ray tracing method simulates thermal radiation by dividing each surface into small elements that emit rays across different angles. Each ray is tracked as it reflects or is absorbed depending on the optical properties of the surfaces it encounters, and reflections continue until all energy is absorbed. By summing these contributions, the method captures the full energy exchange paths, including multiple reflections, enabling accurate analysis of rectification and circulation in the system. Geometrically similar systems with different absolute sizes should yield the same rectification and circulation performance at same temperatures, thought the magnitude of the heat flows can be different. We detail the calculation procedure in the Supplemental Material.

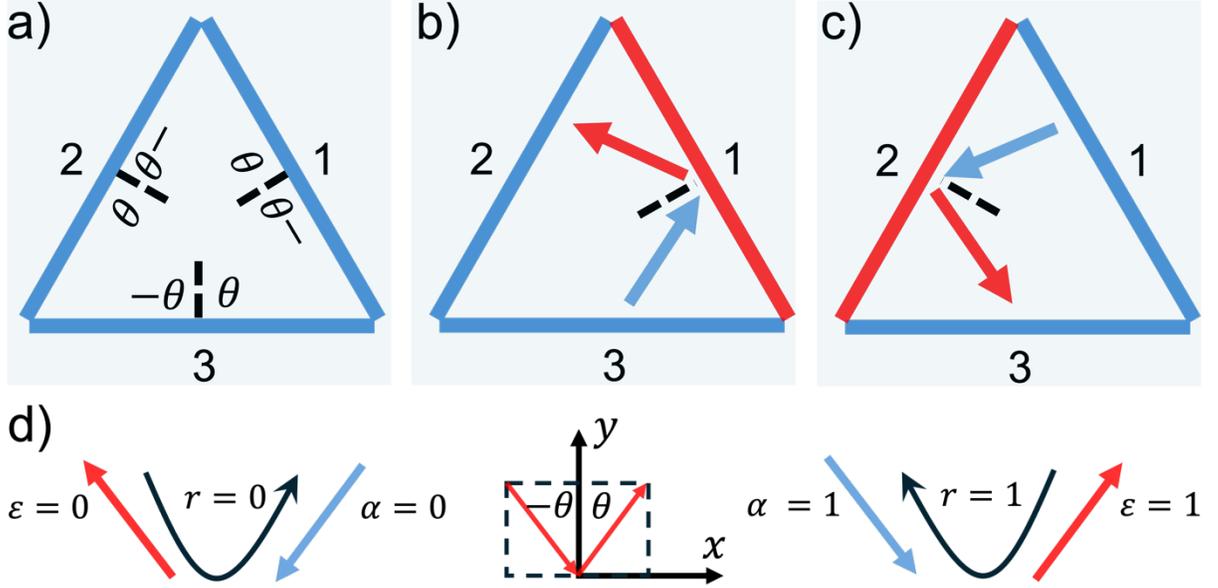

FIG. 2. (a) Schematic of the system and the properties of nonreciprocal surfaces. (b) A representative photon path for the forward (when $T_1 = T_h$) and (c) backward (when $T_2 = T_h$) case. The red arrow indicates the emission from the hot surface, whereas the blue arrow shows the photon received by the hot surface. (d) The angular nonreciprocal radiative properties of each surface. $r(\theta \rightarrow -\theta) = 1$ for light incident from positive polar angles and $r(-\theta \rightarrow \theta) = 0$ for light incident from negative polar angles. Based on Eq. (8), therefore, for $0 \leq \theta \leq 90°$, the absorptivity is zero and the emissivity is one, and vice versa for the $-\theta$ range.

## Results and Discussion

With the nonreciprocal radiative properties, the rectification between surfaces is conceivable. We use surfaces 1 and 2 as an example. As illustrated in Fig. 2(b), when surface 1 is hot, the emission will be directed to surface 2 due to the radiative properties of the surfaces. These rays are mostly absorbed by the surface 2. However, in the backward case when surface 2 is hot, as shown in Fig. 2(c), photons emitted from 2 will be mostly intersected and get absorbed by surface 3, not reaching surface 1. The emission from surface 3 to 1, when nonzero, contributes to the equilibrium persistent heat flow. The asymmetric photon path indicates a rectification between surfaces 1 and 2. With the nonreciprocal ray tracing technique, we compute the interactions of a ray emitted from a given location within the enclosure. By integrating all the emitting points, we obtain the net heat rate between the emitting surface and the rest of the surfaces in the enclosure. In Fig. 3, we show the forward heat transfer rate $q_{f,1}$ (i.e., net heat flow from 1 to 2 when $T_1 = T_h$ and $T_2 = T_3 = T_c$) and backward heat transfer rate $q_{b,2}$ (i.e., net heat flow from 2 to 1 when $T_2 = T_h$ and $T_1 = T_3 = T_c$ ). To simplify our calculation, we fix $T_c$ at 0 K and vary $T_h$ from 300 to 1000 K. We note that the choice of $T_c$ can impact the magnitude of the equilibrium and nonequilibrium heat flow as indicated in Eq. (4), but the rectification is independent of the choice of $T_c$ since it is caused by the nonreciprocity of the system. As shown in Fig. 3, the forward and backward heat flow show a significant difference as indicated by the high rectification ratio $\gamma_{12} = 0.98$. In stark contrast to radiative rectification relying on nonlinear responses triggered by the temperature field[49], $\gamma$ is temperature independent because the wavelength and temperature independent nonreciprocal radiative properties. In practice, the nonreciprocal properties could indeed show strong robustness against temperature change[50]. Thus, the rectification achieved by nonreciprocal radiative properties can potentially be used to regulate photon flow driven by different kinds of thermodynamic forces, such as chemical potential[43].

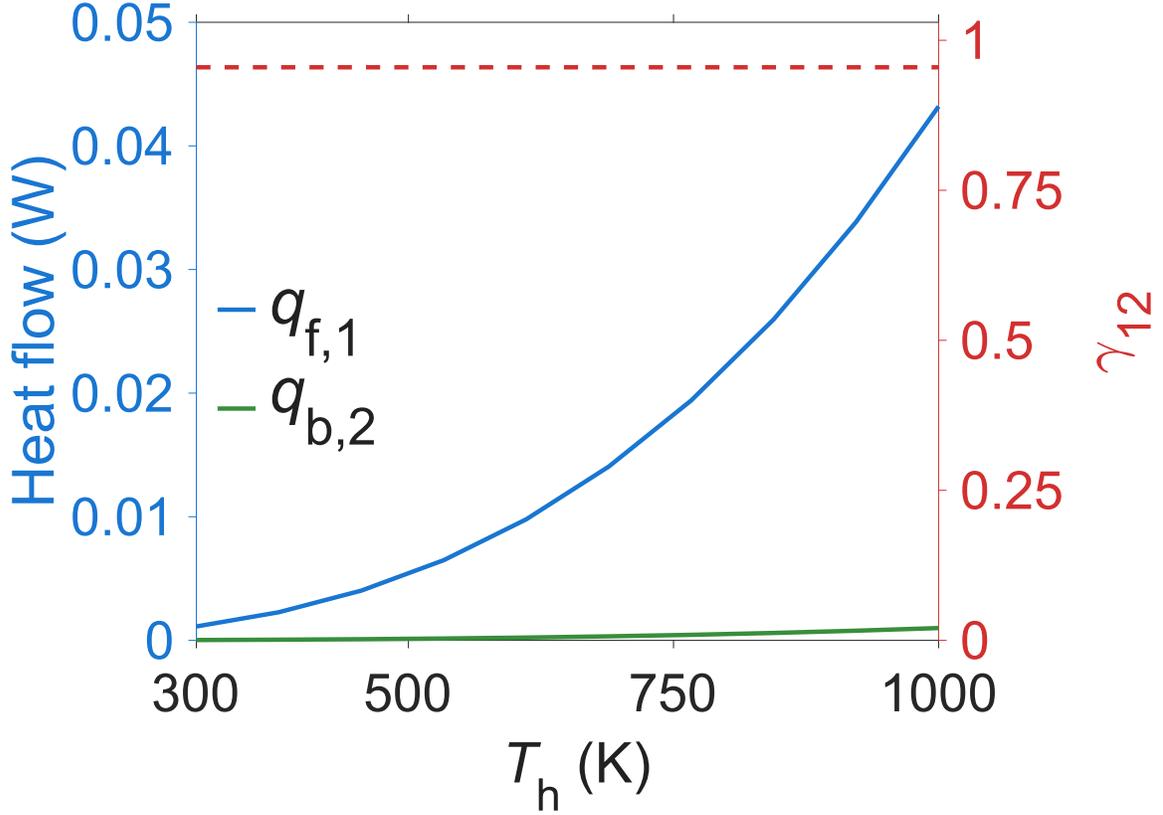

FIG. 3. Heat flow and the rectification ratio between surfaces 1 and 2 in the system shown in Fig. 2. $T_c$ is fixed at 0 K.

It is very interesting to note that, despite the radiative properties in all directions being nonreciprocal, a perfect rectification is not achieved in the system shown in Fig. 2. This implies that $G_{2\to 1}(\omega)$ is not zero, and some photons can still reach surface 1. A closer examination and doing several simulations for different angular range reveal that photons emitted from 2 within the angular range $70° < \theta < 90°$ can indeed reach surface 1 through multiple reflections within the enclosure. We showcase an exemplary photon pathway in the Supplemental Material. Therefore, despite the omnidirectional rectification effect, perfect rectification is still not achieved with the proposed setup and radiative properties.

Based on the insights provided from the ray tracking analysis, we propose to use a modified directional radiative property for the angular range $|\theta| > 70°$. Our goal is to ensure all rays from surface 2 do not be absorbed by surface 1, i.e., perfectly blocking all the possible light paths. In specific, we modified the radiative properties so that all the surfaces have $r = 1$ for $|\theta| > 70°$ as illustrated in Fig. 4(a). In doing so, the rays from surface 2 that can previously reach surface 1 are able to alter their trajectory and being blocked, thereby suppressing any backward heat flow completely and yielding $G_{2\to 1} = 0$. The heat emitted from surface 2 instead is directed to and fully absorbed by surface 3. The system with the modified radiative properties can successfully achieve $\gamma_{12} = 1$, as shown in Fig. 4(b). This shows the importance of using angular-selective emitter and its significant contribution in achieving perfect heat rectification. We note that essentially our goal is to achieving perfect heat rectification, and by modifying the directional properties of the emitter, we are blocking all the pathways that a photon can reach from surface 2 to surface 1. Using angular-selective emitters is one of the approaches to achieve perfect rectification, while there can be other methods to be used to reach the same functionality, including changing the geometry of the enclosure to change the direction of the photons travelling through different surfaces, which is studied in the Supplemental Materials.

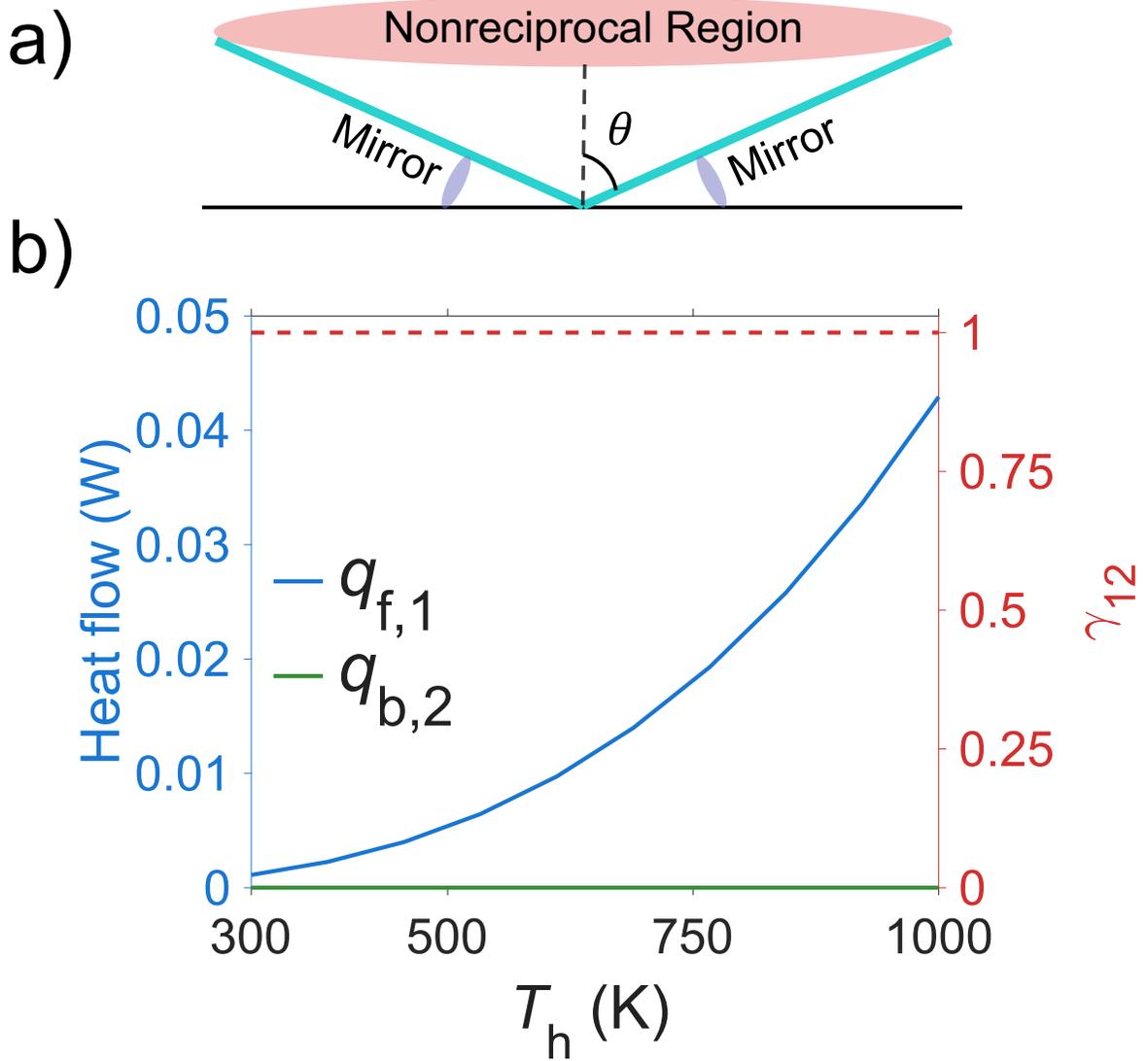

FIG. 4. (a) Schematic showing the angular ranges in which the emitter possesses the perfect nonreciprocal radiative properties depicted in Fig. 2(b) in the polar angle range $|\theta| \leq 70°$ and a specular mirror reflection with $r = 1$ in the polar angle range $|\theta| > 70°$. (b) Heat flow and rectification ratio between surfaces 1 and 2 in the system with the emitters with properties in (a). $T_c$ is fixed at 0 K.

With the perfect rectification between two surfaces, the systems in Fig. 2(a) can naturally achieve perfect circulation for its $C_3$ symmetry. We define a circulation coefficient to quantify the circulation performance in a way that is similar to the rectification:

$$C = \frac{|q_\circlearrowleft - q_\circlearrowright|}{q_\circlearrowleft + q_\circlearrowright}, \tag{9}$$

where $q_\circlearrowright = \sum_{i=1}^{N} q_{f,i}$ and $q_{f,i}$, is the net total heat transfer rate in the forward direction from body $i$ to its subsequential body $j$ by the nonequilibrium term when $T_i = T_h$ and the rest of the system is at $T_c$. When $i = N$, $q_{f,i}$ is the heat flow from body $N$ to body 1. Similarly, $q_\circlearrowleft = \sum_{j=1}^{N} q_{b,j}$ and $q_{b,j}$, is the net total heat transfer rate in the backward direction from body $j$ to its previous body by the nonequilibrium term when $T_j = T_h$ and the rest of the system is at $T_c$. When $j = 1$, $q_{f,j}$ is the heat flow from body 1 to body $N$. In contrast to the figure of merit based on the multiplication of heat flows proposed in existing literature [51-53], here we use the summation so that $C = 1$ can only be reached if $\gamma = 1$ is achieved between each pair of adjacent bodies. Based on Eq. (9), for the system shown in Fig. 2(a), $C = 1$ can be achieved when all surfaces have the radiative properties depicted in Fig. 4(a) that allow perfect rectification.

Besides achieving perfect circulation, the heat that is circulated among different bodies is of the same magnitude due to the $C_3$ symmetry of the system, which makes the circulation ratio also the same as the rectification ratio. The $C_3$ symmetry, however, is not a necessary condition for perfect circulation. In the Supplemental Materials, we provide a detailed study for systems without $C_3$ symmetry. In general, the angular radiative properties and the geometry of the system are coupled and need to be controlled simultaneously to achieve perfect rectification and circulation. For example, our results show that for three surfaces arranged as an obtuse triangle, perfect rectification and circulation can be achieved even for surfaces with omnidirectional nonreciprocal properties shown in Fig. 2. Therefore, it is possible to not only achieve perfect circulation but also control the arrangement of emitters to adapt to different applications. Such freedom can be critical for the realization of the circulators in architectures that can achieve the Landsberg limit[13,54,55] for solar energy harvesting.

We note that the above perfect circulation system naturally supports persistent heat current at thermal equilibrium. Therefore, persistent heat current is a necessary condition of perfect circulation. In the Supplemental Material, we provide a case study to further highlight the difference between persistent heat current phenomena and perfect heat circulation. Persistent heat current is the heat transfer associated with thermal equilibrium terms within the system, whereas heat circulation is the heat transfer with ambient outside the system out of equilibrium. Persistent heat current can exist even when only one of the bodies in the system possesses nonreciprocal properties. It is, however, not sufficient to achieve perfect circulation, which requires much more stringent radiative properties and system configuration.

The previous discussions are focused on systems with ideal nonreciprocal properties that are wavelength independent. In practice, the nonreciprocal properties could show wavelength and polarization dependence, which could result in temperature-dependent rectification and circulation performance. To highlight these impacts, we illustrate a rectification system with one of the state-of-the-art broadband nonreciprocal thermal emitter designs proposed in Ref.[56], as shown in Fig. 5(a). The design is a multilayer structure consisting of three layers of magnetooptical material with increasing doping levels ($n_i$) and three layers of magnetic Weyl semimetals with increasing Fermi levels ($E_{F,i}$). Both $n_i$ and $E_{F,i}$ increases from the top to bottom. The layers are on a silver (Ag) substrate. Each layer provides an epsilon-near-zero region that greatly enhances the nonreciprocity, forming a broad band from 5 - 40 μm within which the nonreciprocity is quite strong and omnidirectional. In the supplemental material, we show the contour plots of the emissivity and absorptivity of the emitter at different wavelengths and angles. The difference between emissivity and absorptivity for such emitter ($\eta$) is plotted in Fig. 5(b) for unpolarized waves. We note that for such structure, nonreciprocal effect only exists for transverse magnetic waves[56] and for transverse electric waves the radiative properties remains reciprocal. So $\eta$ in Fig. 5(b) is all contributed by transverse magnetic waves. Using the same ray tracking technique, we compute the rectification and circulation performance for a system with these surfaces. In these computations, we consider unpolarized emission. We trace the light emitted from a surface until its intensity reaches 1% of its initial value, at which point we stop tracing the beam.

Since the nonreciprocal properties are omnidirectional, we consider an arrangement shown in Fig. 5(d) with the emitters forming an isosceles triangle with surfaces 2 and 3 having the same length and $\beta = 120°$. Such arrangement allows higher rectification ratio for omnidirectional nonreciprocal emitters. As discussed in the Supplemental Material, in the ideal case, the view factor between surfaces 2 and 3 decreases as $\beta$ increases which results in an increasing the rectification ratio. However, the magnitude of the heat flow between these surfaces also decreases. Therefore, we choose $\beta = 120°$, which we can yield a balance between rectification ratio and magnitude of the heat flow.

Figure 5(c) shows the spectral heat transfer rate between bodies $2 \rightarrow 3$ and $3 \rightarrow 2$ for forward and backward case at $T_h = 1000$ K and $T_c = 0$ K. Figure 5(d) shows the rectification ratio and the heat flow between surfaces 2 and 3. Despite that the properties of the emitters are not ideal, decent rectification performance can already be obtained. The strong rectification can persist over a broad temperature range. As the nonreciprocal properties of the emitter are mainly over the mid-infrared range, the rectification ratio is high around room temperature and gradually decrease as the temperature increases. This highlights the importance of the spectral nonreciprocal properties on the rectification and circulation at different temperatures, revealing the necessity for future research to further improve the bandwidth of nonreciprocal thermal radiation for both polarizations for strong rectification and circulation performance.

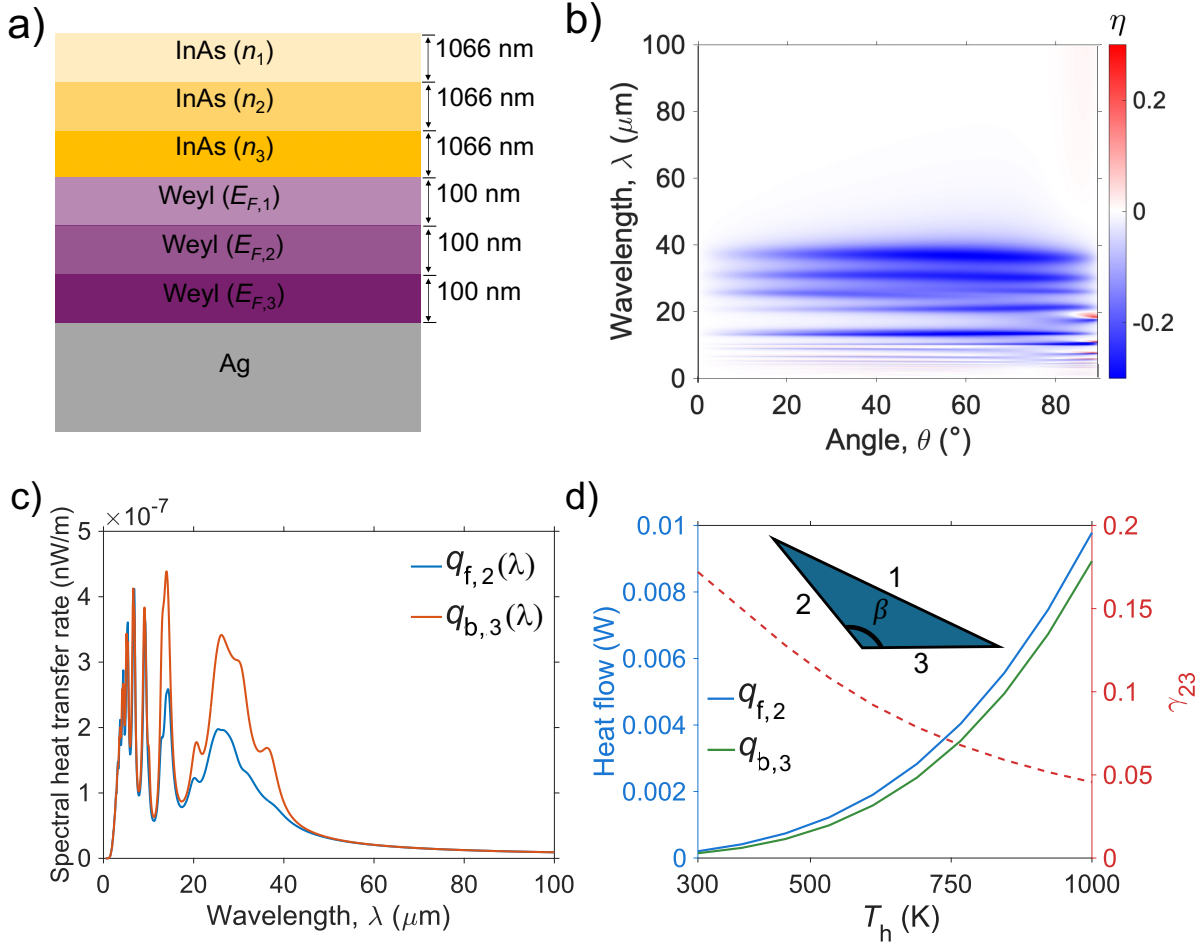

FIG. 5. (a) The emitter proposed in Ref. [56] which provides the state-of-the-art broadband nonpractical behavior. (b) The contrast between absorptivity and emissivity for the emitter shown in (a). (c) The spectral heat transfer rate between surfaces 2 and 3 in the triangle shown on (d), where all surfaces of the triangle are the emitter shown in (a). (d) Heat flow and rectification ratio between surfaces 2 and 3 in a triangle with $\beta = 120°$. All three surfaces have nonreciprocal radiative properties from the design in Ref.[56] . $T_c$ is fixed at 0 K.

## Conclusion

In summary, we have shown that perfect radiative heat rectification and circulation are possible in many-body nonreciprocal systems. When the system is out of equilibrium, the total heat flow among the bodies consists of an equilibrium contribution and a nonequilibrium contribution. While the equilibrium component always exists and is responsible for the persistent heat current, the nonequilibrium component enables the energy exchange between the system and the external environment. With the codesign of angular-selective nonreciprocal radiative properties and geometries, perfect rectifications and circulations can be achieved. Our discoveries put forth a different pathway as compared to nonlinear and time-modulation approaches to achieve high-performance heat rectification and circulation. Future research is also needed to realize the angular-selective broadband nonreciprocal thermal emitters required for perfect rectification and circulation performance.

The authors acknowledge the funding from NSF under Grants No. CBET-2314210 and CBET-2440814, and the support of the Research Computing Data Core at the University of Houston for assistance with the calculations carried out in this work.

**Data availability**

The data used to produce the results within this work are available upon requests to the authors.

## 1. Calculation Methodology

The ray tracing techniques applies to both three-dimensional (3D) and two-dimensional (2D) domain. Here we consider photon transport in the 2D space. The spectral energy density, $u_\nu$, (energy per unit area per unit frequency interval) reads[S1]:

$$u_\nu(\nu)d\nu = \frac{2\pi h \nu^2}{c^2\left(e^{\frac{h\nu}{k_B T}}-1\right)}d\nu. \qquad (10)$$

For a length element inside the enclosure, the spectral emissive power from a blackbody is related to the energy density and the speed of light by:

$$e_{b,\nu}(\nu,T) = \frac{u_\nu c}{\pi} = \frac{2h\nu^2}{c\left(e^{\frac{h\nu}{k_B T}}-1\right)}. \qquad (11)$$

The radiation intensity for a 2D blackbody is then

$$I_{b,\lambda}(\lambda,T) = \frac{e_{b,\lambda}(\lambda,T)}{2}. \qquad (12)$$

Our calculation approach begins by discretizing each surface into a specific number of elements. For each element, we discretize the angular range, and, at each angle, a ray is emitted and propagates within the system. Upon interacting with the surfaces in the system, the energy of the ray will be absorbed or reflected based on the radiative properties of the surface. The ray-tracing algorithm follows each emitted ray until its energy is fully absorbed by the surfaces in the system.

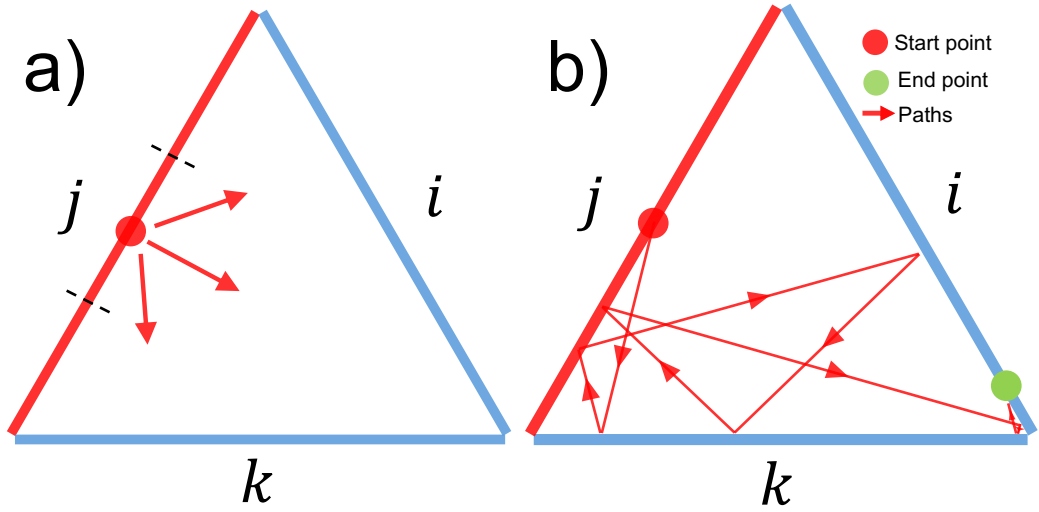

FIG. 6. Schematic for the ray-tracing method. (a) Discretization of each surface into equal segments. (b) The trajectory of the beam is controlled by the optical properties of the interacting surface. The figure illustrates the case mentioned in the main text, where the emitted ray from surface 2 in the angular range $70° < \theta < 90°$ can reach surface 1 through multiple reflections when all three surfaces possess the ideal nonreciprocal properties for all angles. Here, the notations $i, j$, and $k$ correspond to 1, 2, and 3, respectively, as used in the main text for naming each side of the triangle.

As shown in Fig. 6(a), we examine an exemplary triangle by dividing side $j$ into specific number of segments $N_l$, with each segment emitting specific number of rays, $N_a$, in distinct directions. We note that the number of surface elements and the angular range can be different. To calculate the emitted energy from side $j$, we compute the emission at each area element $l$. For each angular element $a$, the emitted energy is given by:

$$Q_{j,l,a} = \varepsilon_j(\lambda, \theta) I_{b,\lambda}(\lambda, T) \cos(\theta). \tag{13}$$

The total energy emitted by each area element is then integrated over the angular range:

$$Q_{j,l} = \sum Q_{j,l,a} \frac{\pi}{N_a}. \tag{14}$$

The total energy emitted by surface $j$ becomes:

$$Q_j = \sum Q_{j,l} \frac{L_j}{N_l}. \tag{15}$$

Here $L$ stands for the length of each side. To calculate the absorbed energy by each side, we trace each emitted ray until it is fully absorbed.

We illustrate an example in Figure 6(b). We follow an emitted ray from surface $j$ that finally reaches surface $i$. Our algorithm identifies the angles at which the ray intersects with different sides while reflecting. Knowing the spectral-directional absorptivity for each angle allows the calculation of the absorbed energy. The initial emitted ray in Fig. 6(b) reaches to surface $k$ first where the absorbed energy is:

$$Q_{abs,k} = \alpha_k(\lambda, \theta) Q_{j,l,a}; \tag{16}$$

The remaining energy is reflected:

$$Q_{ref,k} = r_k(\lambda, \theta) Q_{j,l,a}. \tag{17}$$

The reflected rays continue travelling to other surfaces as per the algorithm's directional reflectivity calculations until get fully absorbed. The total absorbed energy for side $i$ from side $j$ is:

$$Q_{abs,i,total} = \sum Q_{abs,i} \frac{L_j}{N_l} \frac{\pi}{N_a}. \tag{18}$$

## 2. Heat rectification and circulation for general structures

In the main text, we primarily focus on a case of a system with $C_3$ symmetry, as shown in Fig. 7(a). Here, we extend our analysis to general structures that do not have $C_3$ symmetry by keeping the length of sides 2 and 3 the same and varying the angle $\beta$, as shown in Fig. 7(b).

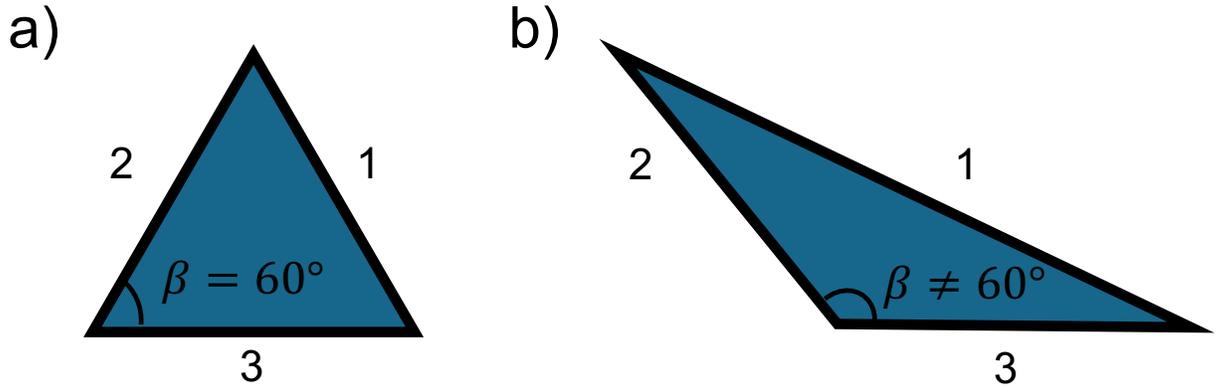

FIG. 7. (a) The system having $C_3$ symmetry and studied in the main text. (b) Altering the geometry of the system and combination of the emitter for each side to break the $C_3$ symmetry. We keep the size of the sides 2 and 3 equal and change the $\beta$.

We consider surfaces with ideal omnidirectional nonreciprocal properties as described in Fig. 2(d) in the main text. As discussed in the main text, the rectification between surfaces 1 and 2 cannot reach perfect with $\beta = 60°$. We illustrate the heat rectification and circulation for three other $\beta$ in Fig. 8. When $\beta = 45°$, the rectification between surfaces 1 and 2 further deteriorates since the pathways depicted in Fig. 6(b) become easier to access. However, as demonstrated in Figs. 8(e)-(l), perfect rectification and circulation become possible when $\beta$ is equal or larger than 90°, indicating a successful suppression of the pathways shown in Fig. 6(b). Therefore, it becomes apparent that $C_3$ symmetry is not a necessary condition for perfect circulation. The required nonreciprocal properties for perfect rectification and circulation are coupled with the geometry of the system. Besides tunning the directional nonreciprocal properties, reconfiguring the geometry of the system can be also utilized to achieve perfect rectification and circulation. The thermal performance of the system can be effectively reconfigured through control of geometries.

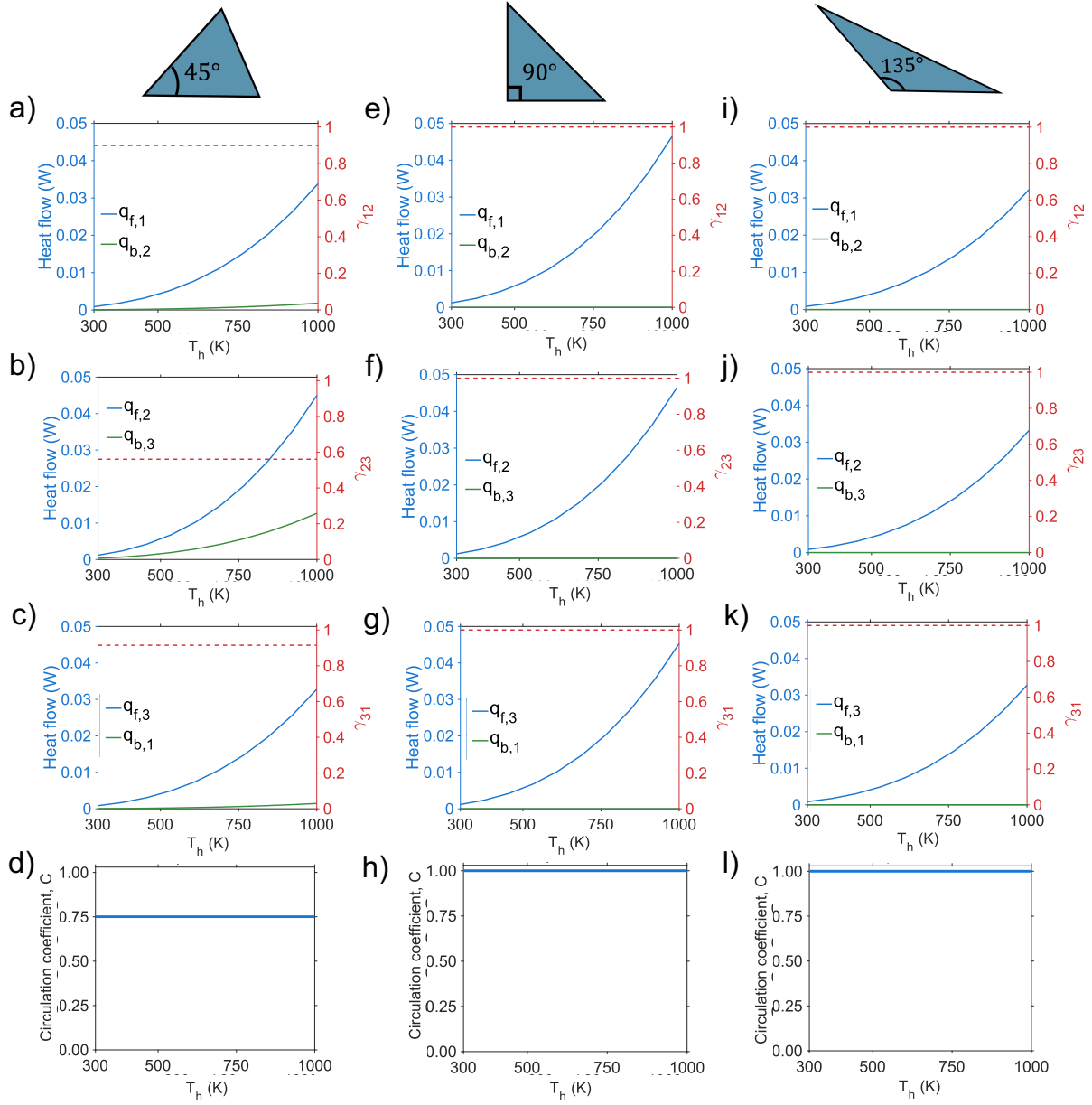

FIG. 8. Heat flow rate, rectification ratio, and heat circulation performance for different geometries when sides 1, 2, and 3 are omnidirectional nonreciprocal emitters. (a-c) Heat flow between sides 1-2, sides 2-3, and sides 3-1, respectively, and (d) heat circulation performance for $\beta = 45°$. e-g) Heat flow between sides 1-2, sides 2-3, and sides 3-1, respectively, and (h) heat circulation performance for $\beta = 90°$. (i-k) Heat flow between sides 1-2, sides 2-3, and sides 3-1, respectively, and (l) heat circulation performance for $\beta = 135°$. $T_c$ is fixed at 0 K.

### 3. Persistent heat current and perfect heat circulation

As mentioned in the main text, a persistent heat current can be achieved even with just one nonreciprocal surface. Here, we demonstrate a specific case to support this claim and highlight the distinctions between persistent heat current and perfect heat circulation. Let us assume that only side 3 in Fig. 7(a) is an omnidirectional nonreciprocal surface, while the other sides are blackbody emitters. Figures 9(a–c) illustrate the heat flow and the rectification ratio between each pair of surfaces. The calculation process is the same as the rectification calculations detailed in the main text. With the calculations, one can derive the performance of the system at equilibrium conditions when all three surfaces are at the same temperature. As an example, we can analyze the heat flow values at 1000 K:

When $T_1$ = 1000 K and $T_2 = T_3$ = 0 K, $q_{1 \to 2}$ = 0.082 W; when $T_2$ = 1000 K and $T_1 = T_3$ = 0 K, $q_{2 \to 1}$ = 0.046 W. Therefore, when $T_1 = T_2$ = 1000 K, the net heat flow from side 1 to side 2 is 0.036 W. We note that this quantity is independent of the choice of $T_3$.

Similarly, from $q_{2 \to 3}$ = 0.042 W, $q_{3 \to 2}$ = 0.006 W, we see that when $T_2 = T_3$ = 1000 K, net heat flow from side 2 to side 3: 0.036 W.

Lastly, from $q_{3 \to 1}$ = 0.042 W, $q_{1 \to 3}$ = 0.006 W, we see that when $T_3 = T_1$ = 1000 K, net heat flow from side 3 to side 1: 0.036 W.

Therefore, we can see that when $T_1 = T_2 = T_3$ = 1000 K, there is a circulating heat current of the magnitude of 0.036 W among the three bodies, forming a directional persistent heat current. We also conducted a control calculation with surface 3 being a reciprocal surface, and no persistent heat current is observed.

In Fig. 9(d), we plot the circulation performance for the same system but under non-equilibrium conditions. As shown, the system exhibits heat circulation with $C$ = 0.48, far from perfect circulation. The results highlight that persistent heat current is only a necessary but not a sufficient condition for perfect heat circulation. Achieving perfect heat circulation demands more stringent system requirements.

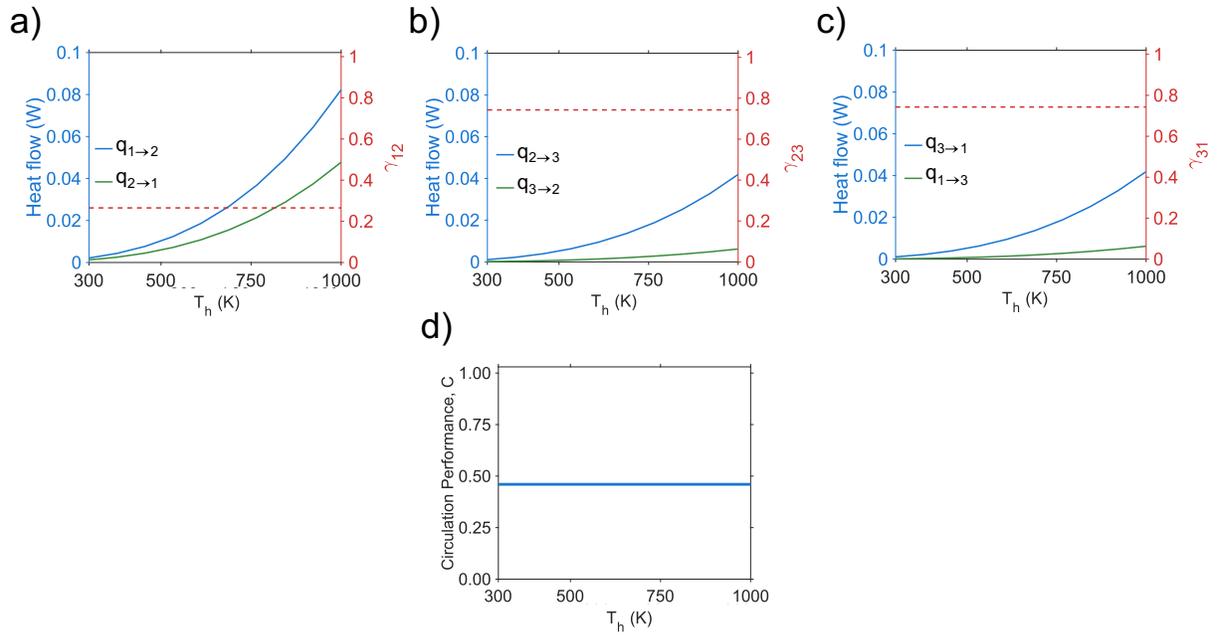

FIG. 9. (a)-(c) Heat flow and rectification ratio for each pair of a system where side 3 is an omnidirectional nonreciprocal emitter and sides 1 and 2 are blackbody emitters. (d) The circulation performance for such a system. $T_c$ is fixed at 0 K.

### 4. The nonreciprocal properties of the current state-of-the-art emitter

As shown in the main text, heat rectification and heat circulation for a system with current state-of-the-art emitter, the design proposed in Ref.[S2], are calculated. This emitter exhibits nonreciprocal behavior in the mid- and far-infrared wavelength ranges, where the spectral intensity of the radiation in the studied temperatures is dominant. Therefore, the proposed emitter in Ref.[S2] is a suitable option for getting strong heat rectification. The emissivity and absorptivity for such emitter is plotted in Figs. 10 (a) and (b) for unpolarized light.

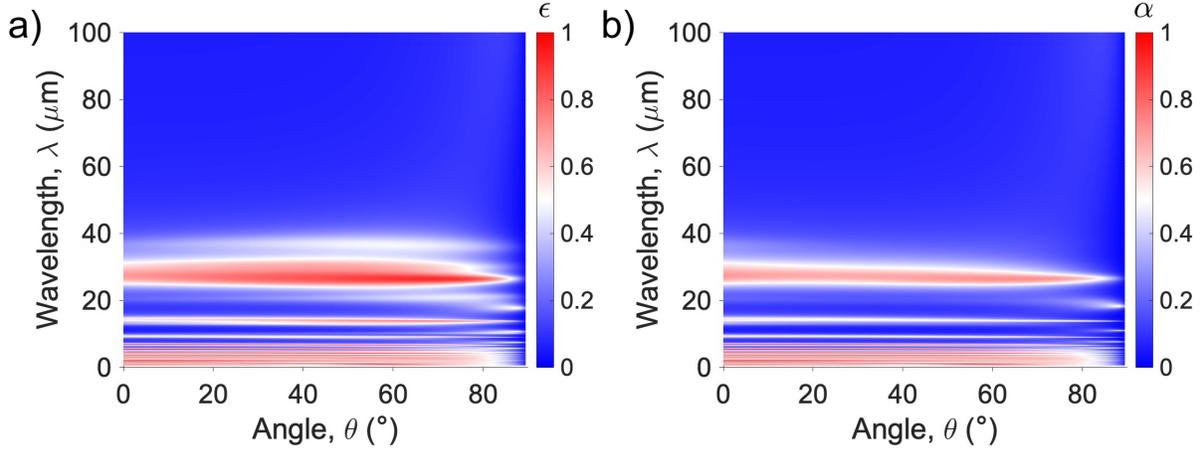

FIG. 10 (a) The contour plot of emissivity, $\varepsilon$, and (b) absorptivity, $\alpha$ for different wavelengths and angles for the emitter proposed in Ref.[S2].

In Fig. 11, the circulation performance (C) of systems using state-of-the-art surfaces is shown for different values of $\beta$. As $\beta$ becomes either very small or very large, the view factor between at least two sides of the triangle decreases. In this case, although heat rectification between one surface pair is very strong, but for the other pairs is weak and leading to reduced circulation performance. In contrast, for $\beta$ that makes the view factor between each pair of the surfaces larger, or in another word each surface can see the other two surfaces more clearly, the circulation performance increases. This highlights the critical role of geometry: strong and balanced heat rectification between each pair of surfaces is required to achieve high circulation performance.

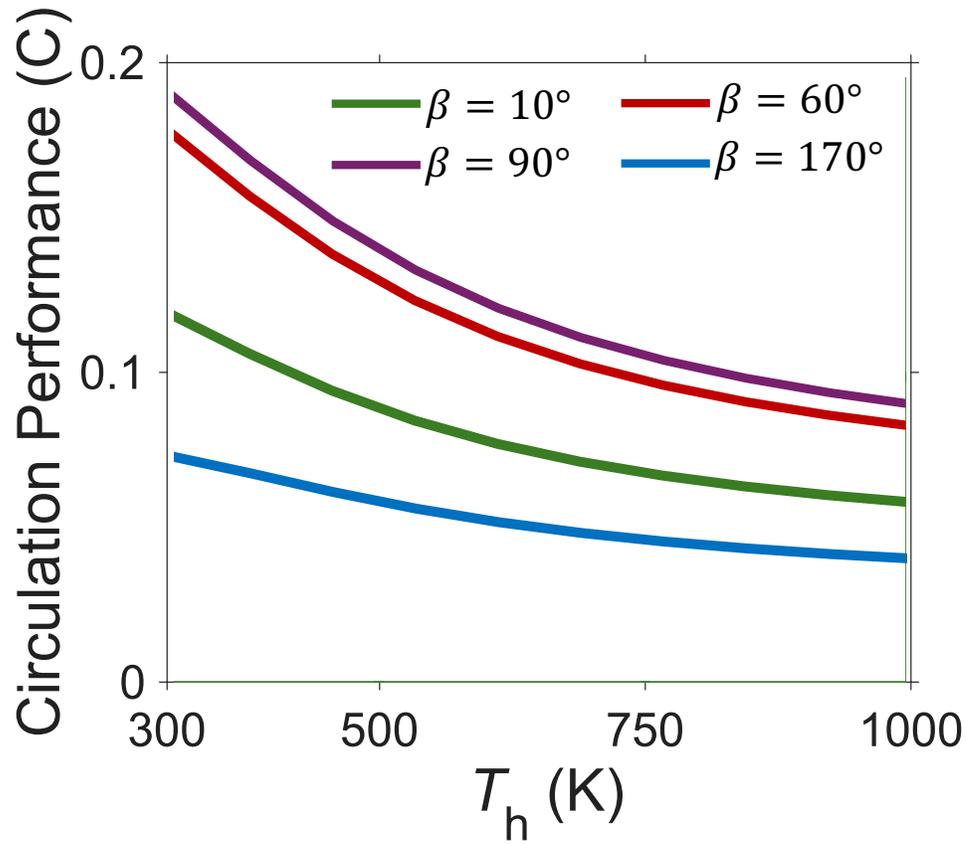

FIG 11. The heat circulation performance for different values of $\beta$ for different temperatures.